\documentclass[11pt, a4paper, onecolumn, accepted=2021-12-16]{quantumarticle}

\pdfoutput = 1

\usepackage{textcase}
\usepackage[square,sort,comma,numbers]{natbib}
\usepackage{braket}
\usepackage{graphicx}
\usepackage{hyperref}
\usepackage{amssymb}
\usepackage[utf8]{inputenc}
\usepackage{float}
\usepackage{doi}
\expandafter\let\csname equation*\endcsname\relax

\expandafter\let\csname endequation*\endcsname\relax

\usepackage{amsmath}

\setcitestyle{square}

\makeatletter
\newcommand*\bigcdot{\mathpalette\bigcdot@{.5}}
\newcommand*\bigcdot@[2]{\mathbin{\vcenter{\hbox{\scalebox{#2}{$\m@th#1\bullet$}}}}}
\makeatother

\begin{document}

\title{Lieb's Theorem and Maximum Entropy Condensates}
\author{J. Tindall}
\email{joseph.tindall@physics.ox.ac.uk}
\affiliation{Clarendon Laboratory, University of Oxford, Parks Road, Oxford OX1 3PU, United Kingdom}
\author{F. Schlawin}
\affiliation{Max Planck Institute for the Structure and Dynamics of Matter, 22761 Hamburg, Germany}
\affiliation{The Hamburg Centre for Ultrafast Imaging, Luruper Chaussee 149, Hamburg, Germany}
\author{M. A. Sentef}
\affiliation{Max Planck Institute for the Structure and Dynamics of Matter, 22761 Hamburg, Germany}
\author{D. Jaksch}
\affiliation{Clarendon Laboratory, University of Oxford, Parks Road, Oxford OX1 3PU, United Kingdom} 
\affiliation{The Hamburg Centre for Ultrafast Imaging, Luruper Chaussee 149, Hamburg, Germany}
\affiliation{Institut für Laserphysik, Universität Hamburg, 22761 Hamburg, Germany}

\begin{abstract}
Coherent driving has established itself as a powerful tool for guiding a many-body quantum
system into a desirable, coherent non-equilibrium state. A thermodynamically large system will, however, almost always saturate to a featureless infinite temperature state under
continuous driving and so the optical manipulation of many-body systems is considered feasible only if a transient, prethermal regime exists, where heating is suppressed. Here we show that, counterintuitively, in a broad class of lattices Floquet heating can actually be an advantageous effect. Specifically, we prove that the maximum entropy steady states which form upon driving the ground state of the Hubbard model on unbalanced bi-partite lattices possess uniform off-diagonal long-range order which remains finite even in the thermodynamic limit. This creation of a `hot' condensate can occur on \textit{any} driven unbalanced lattice and provides an understanding of how heating can, at the macroscopic level, expose and alter the order in a quantum system. We discuss implications for recent experiments observing emergent superconductivity in photoexcited materials.
\end{abstract}

\maketitle


\section{Introduction}
Lattice geometry plays a decisive role in the properties of discrete quantum systems. In the fermionic Hubbard model - a successful, simple description of correlated fermions in solid-state materials - the lattice structure is known to significantly influence the system's equilibrium phases. The presence of triangles in the underlying lattice, for example, causes frustration, which prevents regular antiferromagnetic ordering \cite{HubbardFrustration1, HubbardFrustration2, HubbardFrustration3} and induces the formation of exotic phases of matter such as a spin-liquid \cite{HubbardSpinLiquid1, HubbardSpinLiquid2}.   
\par Meanwhile, in the context of repulsive bi-partite Hubbard models, Lieb showed there is a strong distinction between the ground states on \textit{balanced} vs \textit{unbalanced} lattices \cite{LiebTheorem} - with the former meaning the number of sites in each of the two sublattices are equal and the latter meaning they are not. For the balanced case, the ground state has a total spin of zero and is thus an antiferromagnet \cite{QuasiLongRange1}. In the unbalanced case Lieb proved that the total spin is finite, resulting in a ferrimagnetic ground state \cite{FerrimagneticHubbard1, FerrimagneticHubbard2, FerrimagneticHubbard3}.
\par Whilst the equilibrium properties of the Hubbard model on unbalanced lattices are well known, the system's response to external forces such as a periodic driving field is not. This is especially true in comparison to the extensive theoretical efforts on balanced, hypercubic lattices to engineer driven Hamiltonians which guide the system into ordered, prethermal phases whilst transiently mitigating the deleterious Floquet heating \cite{SCTheory1, SCTheory2, SCTheory3, SCTheory4, SCTheory5, SCTheory6} which is almost always inevitable - with only a few exceptions known \cite{FloquetHeatingException1, FloquetHeatingException2}. These efforts to realise such correlated prethermal states of matter are motivated by the opportunities arising from the realisation of coherently driven electronic systems in solid-state material experiments \cite{Tancogne-Dejean_Ultrafast_2018, Ishikawa2014, Wall2011}. 
\par Solid-state materials, however, form a variety of complex, geometrical structures that are typically not balanced, hypercubic lattices \cite{CopperOxideLayers1, CopperOxideLayers2, Katayama2021, 1998cond.mat..2198M}. Moreover, techniques such as scanning tunnelling microscopy and chemical synthesis mean that experimentalists now have an enormous degree of control over the lattice structure of the correlated materials they can synthesise \cite{STM1, STM2, STM3, ChemicalSynthesis1, ChemicalSynthesis2, ChemicalSynthesis3, ChemicalSynthesis4, ChemicalSynthesis5}. Moire superlattices - stacked sheets of Van der Waals heterostructures - are such materials \cite{Moire1, Moire2, Moire3, Moire4}. The tunability of their geometry, physical dimension and frustration has made them a promising condensed matter alternative to ultracold atomic setups \cite{MoireHubbard1, MoireHubbard2, MoireHubbard3}, where the non-equilibrium physics of the Hubbard Hamiltonian can be directly simulated \cite{UltracoldHubbard1, Messer_Floquet_2018}. We thus consider it both pertinent and timely to determine the differences which arise in the response of correlated electronic systems to external driving fields when moving away from balanced, hypercubic setups and onto more complex lattices. 
\par In this paper we show that the maximum entropy states which form upon continuous Floquet heating of the ground state of the Hubbard model on \textit{unbalanced} bi-partite lattices can possess uniform off-diagonal long-range order (ODLRO) which remains finite and does not decay away with increasing system size. These states are thus distinct from the featureless infinite temperature state which will always form under Floquet heating on \textit{balanced} bi-partite lattices in the thermodynamic limit. We prove this result by applying Lieb's theorems in conjunction with the constraints imposed by periodic driving which preserves the relevant SU(2) symmetry. These constraints, in tandem with the heating induced by the driving, force a dynamical renormalisation of the `staggered', long-range correlations initially present in the ground state - fundamentally altering them and the type of order present in the system. The resulting maximum entropy\footnote{Naturally, in this work we mean `maximum entropy' in the relevant SU(2) subspace, as opposed to in the full Hilbert space of the system. This is what allows such states to possess non-zero off-diagonal order.} state is completely translationally invariant and possesses ODLRO.
\par Using recently developed analytical methods we quantify this order for all possible bi-partite lattices in the limit of a large number of sites. In the attractive regime on any unbalanced lattice the steady state is arguably a superconductor as the ODLRO is in the charge sector and the Meissner effect and flux quantisation are manifest. The same cannot be said for the ground state as it lacks translational invariance. We then detail how these ordered, maximum entropy steady states can be realised and probed with current technology in ultracold atomic simulators. In this setting properties such as the optical conductivity spectrum can be determined, which is not possible with current computational methods. 
\par Our results provide an understanding, independent of any microscopic parameters, of how strong driving can expose and rearrange the order in a quantum system at the macroscopic level. We conclude by discussing the possible connection between this phenomenon and observations of emergent superconductivity in irradiated solid-state materials.
\section{Theory and Results}
Our starting point is the fermionic Hubbard model defined over some lattice $\Lambda$ with $N$ sites. The Hamiltonian reads
\begin{equation}
H = - \sum_{\substack{a,b \ \in \ \Lambda \\ a \neq b}}J_{a,b}\sum_{\sigma}c^{\dagger}_{a, \sigma}c_{b, \sigma} + U\sum_{a \in \Lambda}n_{a, \uparrow}n_{a, \downarrow},
\label{Eq:Hamiltonian}
\end{equation}
where $n_{\sigma, a}$, $c_{\sigma, a}^{\dagger}$ and $c_{\sigma, a}$ are, respectively, number, creation and annihilation operators for fermions of spin $\sigma \in \{\uparrow, \downarrow\}$ on site $a$ within the lattice. The first summation in $H$ runs over the sites of the lattice and kinetically couples them together with strength $J_{a,b}$; for Hermiticity we have $J_{a,b} = J_{b,a}$. The second term in $H$ represents an interaction of strength $U$ between the two spin species on a given site $a$. 
\par We assume that the lattice $\Lambda$ is bi-partite, i.e the vertices can be split into two sublattices $A$ and $B$ with $J_{a,b} = 0$ if both $a$ and $b$ are in the same sublattice. Without loss of generality we set $N_{A} \geq N_{B}$, where $N_{A}$ and $N_{B}$ are the number of sites in each sublattice. 
We can also assume, without loss of generality, that the system is connected, i.e. each site can be reached from any other by moving between pairs of sites connected by a `bond' (i.e where $J_{a,b} \neq 0$). Furthermore we fix ourselves to half-filling, i.e. $\sum_{a}\langle n_{a, \uparrow} + n_{a, \downarrow} \rangle = N = N_{A} + N_{B}$ and ensure the total number of lattice sites $N$ is even - conditions which are necessary for Lieb's theorem to be valid \cite{LiebTheorem}. Finally, for simplicity, we fix ourselves to zero $z$-magnetisation, i.e. $\sum_{a}\langle n_{a, \uparrow} - n_{a,\downarrow} \rangle = 0$. Our results, however, should also apply at other magnetisations. 

\par The Hamiltonian in Eq. (\ref{Eq:Hamiltonian}) is ${\rm SU(2)} \times {\rm SU(2)}$ symmetric. The first of these SU(2) symmetries, known as the `spin' symmetry, can be introduced through the spin-raising operator $S^{+} = \sum_{a}S^{+}_{a} = \sum_{a}c_{a, \uparrow}^{\dagger}c_{a, \downarrow}$, its conjugate $S^{-}$ and the total $z$-magnetisation $S^{z} = \sum_{a}(n_{\uparrow, a} - n_{\downarrow, a})/2$. These each commute with $H$. The second, known as the `$\eta$' symmetry, can be introduced via the $\eta$-raising operator $\eta^{+} = \sum_{a}\eta^{+}_{a} = \sum_{a}f(a)c_{a, \uparrow}^{\dagger}c_{a, \downarrow}^{\dagger}$, its conjugate  $\eta^{-}$ and the modified total number operator $\eta^{z} = \sum_{a}(n_{\uparrow, a} + n_{\downarrow, a} - 1)/2$. The latter of these commutes directly with $H$ whilst the former commute with $H$ once a trivial constant term is added to the Hamiltonian and the function $f(a)$ is set to be $\pm 1$ depending on whether the vertex $a$ is in $A$ or $B$. Importantly, the two symmetries are fundamentally related as the mapping $U \leftrightarrow -U$ is effectively equivalent to swapping the spin and $\eta$ degrees of freedom, i.e. $\eta \leftrightarrow S$ \cite{HubbardSymmetries}.

\par This transformation, along with the theorems derived by E. Lieb \cite{LiebTheorem}, can be used to show that for $U > 0$ the ground state of $H$ satisfies $\langle S^{2} \rangle = X(X+1)$ and $\langle \eta^{2} \rangle = 0$ with $X$ = $(1/2)(N_{A} - N_{B})$. Meanwhile, for $U < 0$, we have $\langle S^{2} \rangle = 0$ and $\langle \eta^{2} \rangle = X(X+1)$. The operators $S^{2} = (S^{z})^{2} + (1/2)(S^{+}S^{-} + S^{-}S^{+})$ and $\eta^{2} = (\eta^{z})^{2} + (1/2)(\eta^{+}\eta^{-} + \eta^{-}\eta^{+})$  are, respectively, the spin and $\eta$ Casimir operators.

\par This result has significant implications for the ground state properties of $H$. For positive $U$, in \textit{balanced} bi-partite lattices where $N_{A} = N_{B}$ (examples include any hypercubic lattice) the electron spins on sites in identical sublattices tend to align whilst those in separate sublattices misalign, creating an antiferromagnetic state with $\langle S^{2} \rangle = 0$. The distance dependence of these correlations varies with the lattice dimension; along any spin-axis the correlations in a 1D chain are known to decay algebraically with distance \cite{HubbardSymmetries, QuasiLongRange2} whilst in a 2D square lattice the correlations along the $z$ spin-axis are oscillatory and long-range \cite{QuasiLongRange3, QuasiLongRange4}. 
\par Meanwhile, for \textit{unbalanced} lattices, by definition, one of the sublattices is larger than the other and thus the same alignment tendencies force the system into a ferrimagnetic state instead with $\langle S^{2} \rangle > 0$. Moreover, regardless of the physical dimension, for \textit{any} lattice where $N_{A} - N_{B} \propto N$ the correlations along both the $x$ and $y$ spin-axes will not decay, creating a long-range, oscillatory profile with a non-zero average \cite{HubbardLRO, SUNHubbard}. For negative $U$ the same is true, but for the correlations in the charge sector instead of the spin sector.

\begin{figure}[t]
\centering
\includegraphics[width = \columnwidth]{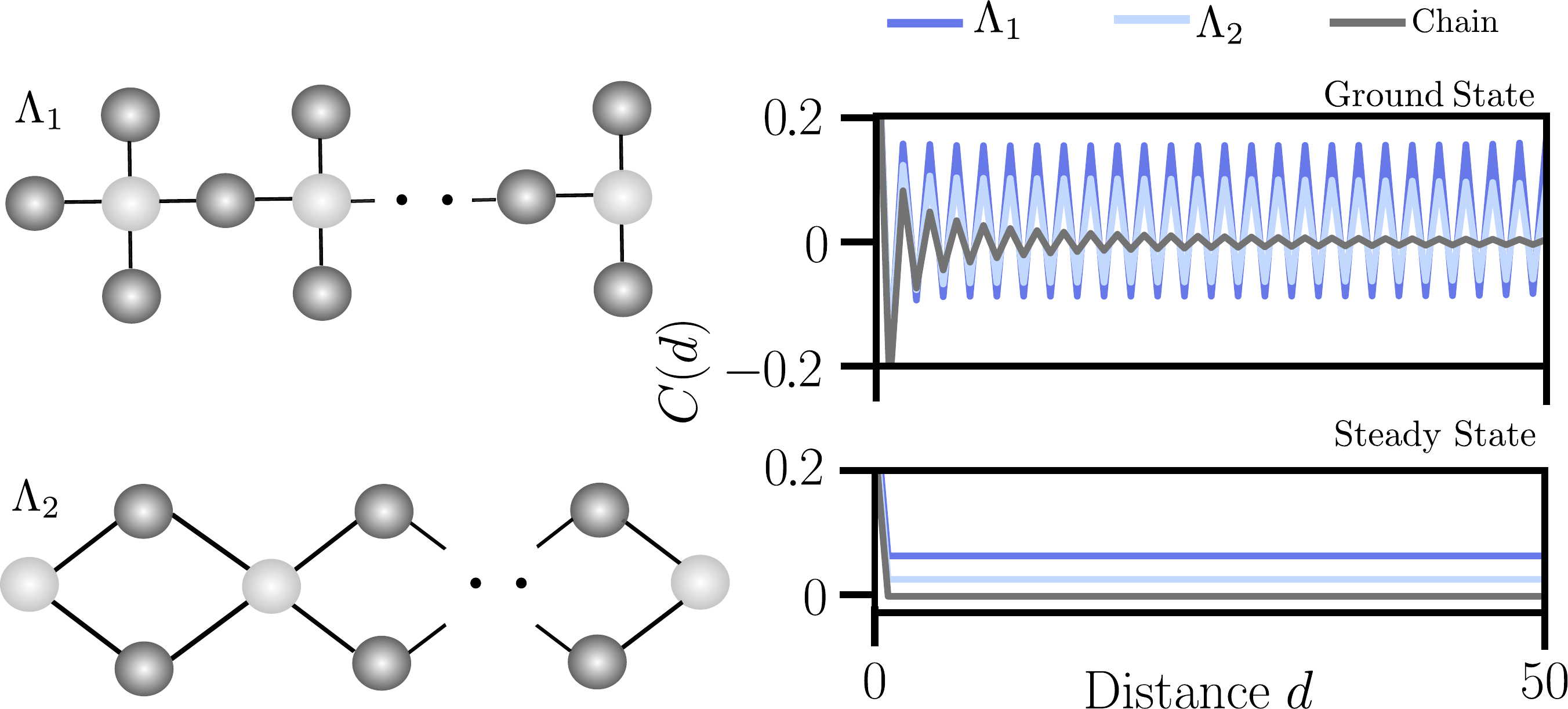}
\caption{Left) Two open boundary unbalanced bi-partite lattices $\Lambda_{1}$ and $\Lambda_{2}$. The light vs dark sites indicate the two sublattices and the bonds correspond to coupled sites where the kinetic coupling $J_{a,b} = J \neq 0$. For all other pairs of sites $J_{a,b} = 0$.  Right) Correlations versus distance for the ground state and maximum entropy steady states of the half-filled Hubbard model on $\Lambda_{1}$ and $\Lambda_{2}$, as well as a 1D chain. Each lattice is taken to have $N=100$ sites and we set $U = \pm 5J$. For $U = 5J$, the correlation function at a given distance $d$ is defined in Eq. (\ref{Eq:Correlation Function}) with $O = S$. For $U = -5J$, $C(d)$ is of the same form but with $O = \eta$. The maximum entropy state is that which forms when the ground state is Floquet heated to infinite temperature whilst the relevant SU(2) symmetry (spin for $U > 0$ and $\eta$ for $U< 0$) is preserved. We have kept $d \leq 50$ so the correlation function for each lattice is on the same scale: $d=50$ is the maximum distance in lattice $\Lambda_{1}$ when $N = 100$.}
\label{Fig:F1}
\end{figure}

\par In Fig. \ref{Fig:F1} we illustrate some of these properties by using DMRG \cite{DMRG} to determine the ground state and its correlation profile for several open-boundary quasi-1D Hubbard lattices with $N = 100$ to minimise any boundary or finite-size effects. To aid our analysis we introduce the distance measure $|a-b|$ which is the minimum number of edges (bonds between pairs of sites with non-zero $J$) that must be traversed to move between sites $a$ and $b$. In graph theory terms, $|a-b|$ is the shortest length path between $a$ and $b$. Using this quantity, we can define the distance-dependent correlation function
\begin{equation}
    C(d) = \frac{1}{\mathcal{N}}\sum_{\substack{a,b \in \Lambda \\ |a - b| = d}}\langle O^{+}_{a}O^{-}_{b} \rangle,
    \label{Eq:Correlation Function}
\end{equation}
where $O$ is either $S$ or $\eta$ depending on whether U is positive or negative and $\mathcal{N}$ is the number of pairs of sites which satisfy $|a-b| = d$. This function thus represents the \textit{average} of the two-point correlations for all pairs of sites in the lattice separated by a distance $d$. Such a function can be used to calculate the distance-dependence of two-point correlations for a many-body system on \textit{any} lattice and allows us to directly compare different lattice structures.
\par For the ground states of the unbalanced lattices in Fig. \ref{Fig:F1} this correlation function is non-decaying and staggered in both sign and magnitude as a function of distance. On the balanced 1D chain this staggering is also present, but decays away with distance. For simplicity, we have taken the coupling strength $J_{a,b}$ to be homogeneous, i.e. $J_{a,b} = {\rm const.}$ if it is non-zero. For inhomogeneous $J_{a,b}$ on the unbalanced lattices the correlations in the ground state will become more disordered but they will still retain a similar, oscillatory, long-range structure as Lieb's theorems are independent of the specific $J_{a,b}$.

\par The central result of this work is that driving the ground states of unbalanced lattices to maximum entropy/ infinite temperature\footnote{In the sense that the driven state is indistinguishable, at the level of expectation values of few-body observables, from the identity matrix over the available SU(2) basis states.} with some arbitrary field which preserves the relevant SU(2) symmetry forces the oscillatory correlations to become completely uniform with distance whilst remaining finite and not decaying to $0$ with increasing system size. 
\par Recent work has shown that, generally, the maximum entropy state which forms in the long-time limit of periodic driving will host completely uniform off-diagonal order in the preserved symmetry channel \cite{HeatingInducedOrder1, HeatingInducedOrder2}. Thus far, when driving an equilibrium state this induced order has only been shown to be finite for finite systems and on regular hypercubic lattices it will asymptotically decay to $0$ as $1/N$ \cite{HeatingInducedOrder3}; preserving the intuition of Floquet heating as a deleterious effect in the thermodynamic limit. Here we show that on unbalanced lattices this will not be the case and periodic driving can be used to reach a maximum entropy condensate which hosts finite, uniform ODLRO.

\par This remarkable result can be proven independently of any specific parameters of the Hamiltonian and applied driving field. Consider $U > 0$ and an arbitrary \textit{unbalanced} Hubbard lattice. Provided both $N_{A}$ and $N_{B}$ grow proportionally with $N$ we have that, following Lieb's theorem, the ground state satisfies $\langle S^{2} \rangle \propto N^{2}$. Under the application of continuous periodic driving which preserves $\langle S^{2} \rangle$ then Floquet heating will occur but the system will be restricted to the subspace spanned by the eigenvectors of $S^{2}$ which possess the same value of $\langle S^{2} \rangle$ as the ground state. The pure steady state $\rho_{\rm ss} = \lim_{t \rightarrow \infty}\ket{\psi(t)}\bra{\psi(t)}$ is indistinguishable (at the level of the expectation values of few-body observables) from the identity matrix in this subspace \cite{LongTimeHeating1, LongTimeHeating2, HeatingInducedOrder3}. This identity matrix inherits the permutational invariance of $S^{2}$ and thus the steady state state must have $\langle S^{+}_{a}S^{-}_{b} \rangle = C_{\rm off-diag} \ \ a \neq b$ as well as $\langle S^{+}_{a}S^{-}_{a} \rangle = C_{\rm diag}$, where both $C_{\rm off-diag}$ and $C_{\rm diag}$ are constant. By definition, it then follows that
\begin{align}
    \langle S^{+}S^{-} \rangle = N C_{\rm diag} + N(N - 1)C_{\rm off-diag}.
\end{align}
and as $C_{\rm diag}$ is a local on-site density we have $0 \leq C_{\rm diag} \leq 1$. Moreover, as we are in the zero magnetisation sector we also have $\langle (S^{z})^{2}\rangle = 0$ and $\langle S^{+}S^{-}\rangle = \langle S^{-}S^{+}\rangle$. Hence, $C_{\rm off-diag}$ must be be non-zero and not decay to $0$ with increasing system size in order for $\langle S^{2} \rangle = \langle S^{+}S^{-}\rangle  \propto N^{2}$ and the maximum entropy state hosts completely uniform ODLRO in the spin sector. If, instead, we have $U < 0$ we can apply the same argument to show that the correlations $\langle \eta^{+}_{a}\eta^{-}_{b} \rangle$ will be completely uniform with distance and remain finite in the limit of large $N$.

\par We now directly quantify the order of these maximum entropy states for different lattices. We utilise a recently constructed complete basis which simultaneously diagonalises both $S^{2}$ and $\eta^{2}$ in the Hubbard model for any filling \cite{HeatingInducedOrder3}. This allows us to make analytical predictions for the size of the diagonal and off-diagonal $\eta$ and spin correlations in the maximum entropy states on any lattice.
\par In Fig. \ref{Fig:F1} we plot the correlation function in Eq. (\ref{Eq:Correlation Function}) for the two pictured unbalanced lattices and the 1D Hubbard chain. Whilst the lattices used are of finite-size, any differences to lattices with a larger $N$ are minimal. We observe that in all cases the steady state possesses completely uniform correlations in the relevant symmetry channel. The chain is distinct from the two unbalanced lattices, however, the steady state correlations are over an order of magnitude less and will become exactly zero as $N \rightarrow \infty$. Unlike the ground state, the steady state correlations are completely independent of the specific values of the $J_{a,b}$ and $U$, only the sign of $U$ is relevant.

\begin{figure}[t]
\centering
\includegraphics[width = \columnwidth]{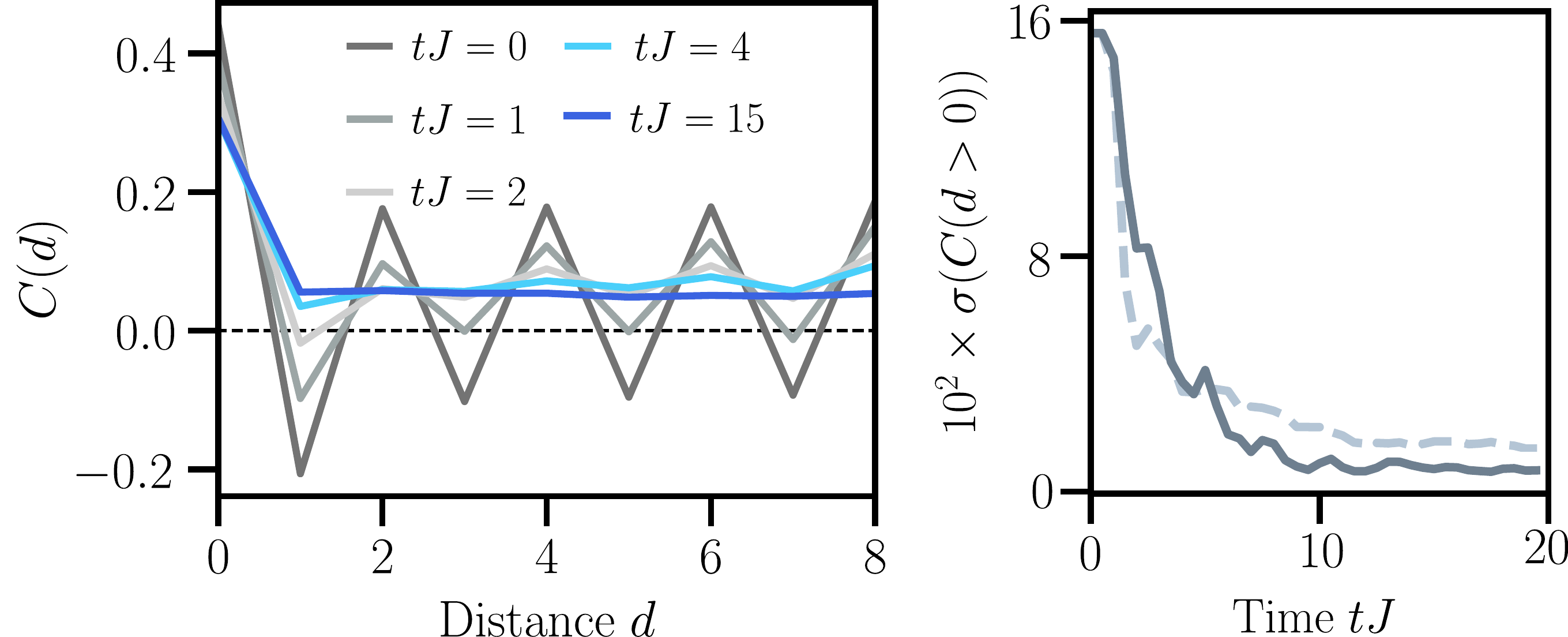}
\caption{Dynamics of the spin correlations in the driven Hubbard model on the saw lattice $\Lambda_{1}$ pictured in Fig. \ref{Fig:F1} with $N=16$ lattice sites. The system is initialised in the ground state with $U = 5J$ and then time-evolved under the same Hamiltonian $H$ but with an additional driving term either of the form $H_{\rm d,1} = A{\rm cos}(\omega t)\sum_{a=1}^{N}a (n_{\uparrow, a}+n_{\downarrow, a})$ with $A = 2J$, $\omega J = 2.5$ (and $a$ - the site index - runs over the lattice sites, increasing from left-to-right and top-to-bottom) or $H_{\rm d, 2} = A{\rm cos}(\omega t)\sum_{a=1}^{N}\epsilon_{a}(n_{\uparrow, a}+n_{\downarrow, a})$ with $A = 10J$, $\omega J = 2.5$ and $\epsilon_{a}$ a random number drawn uniformly on the interval $[-1, 1]$. The time-dependent Hamiltonian is thus $H(t) = H + H_{d,\alpha}(t)$ with $\alpha = 1$ or $\alpha = 2$. Left) Spin correlation $C(d)$ (see Eq. (\ref{Eq:Correlation Function}, we set $O=S$) versus distance $d$ for various instances of time and with the driving term $H_{\rm d, 1}$. Right) Standard deviation of $C(d>0)$ versus time $tJ$. Solid line corresponds to the driving term $H_{\rm d,1}$ and so reflects the data in a) whilst the dashed line corresponds to dynamics with the term $H_{\rm d,2}$. Dynamics in both cases were calculated numerically using the Time-dependent Variational Principle (TDVP) method and a bond dimension of $\chi =8000$.} 
\label{Fig:F15}
\end{figure}

\par Notably, for attractive $U$, the changing sign in the charge order of the ground states is related to the factor of $\pm 1$ used in the $\eta$ operators. If this factor is ignored then the correlations become uniform in sign (but obviously remain inhomogeneous in magnitude) suggesting that the corresponding particle-hole pairs have $Q = 0$ total momentum and the charge order is of the Bardeen-Cooper-Schrieffer (BCS) type. Meanwhile, in the steady state the $\eta$ correlations are homogeneous in both sign and magnitude and so they are of the $\eta$ type with the particle-hole pairs posessing a momentum of $Q=\pi$. In forcing all the correlations to become uniform with distance, the heating has directly changed the type of charge order present in the system.
\par A similar phenomenon also happens for repulsive $U$. Here, the changing sign in the ground state spin order indicates ferrimagnetism, with the electron spins in the two different sublattices being misaligned. In the steady state the constant sign instead indicates ferromagnetism with all spins aligning in the same direction, regardless of the sublattice they are in. The heating has thus dynamically altered the underlying magnetic order in the system.
\par In Fig. \ref{Fig:F15} we demonstrate an explicit example of these dynamics. We consider the ground state of the repulsive Hubbard model on the lattice $\Lambda_{1}$ from Fig. \ref{Fig:F1} with $N=16$ sites and time-evolve it explicitly for two different periodic driving terms which preserve the spin SU(2) symmetry. Fig. \ref{Fig:F15} directly demonstrates the complete re-arrangement undergone in the spin degrees of freedom as the system moves towards the ferromagnetic steady state, with the correlations becoming increasingly uniform in sign and magnitude as a function of distance. We also plot the standard deviation of these correlations as a function of time, directly witnessing this quantity asymptotically decaying to $0$. We use two driving terms to emphasize that provided the spin SU(2) symmetry is preserved this rearrangement will always occur. The specific parameters and driving term only affect the timescale on which it occurs. Moreover, by letting $U \rightarrow -U$ in the Hamiltonian and $n_{\uparrow, a}+n_{\downarrow, a} \rightarrow n_{\uparrow, a}-n_{\downarrow, a}$ in the driving term then the same results will occur for the correlations in the charge degrees of freedom.

\begin{figure}[t]
\centering
\includegraphics[width = \columnwidth]{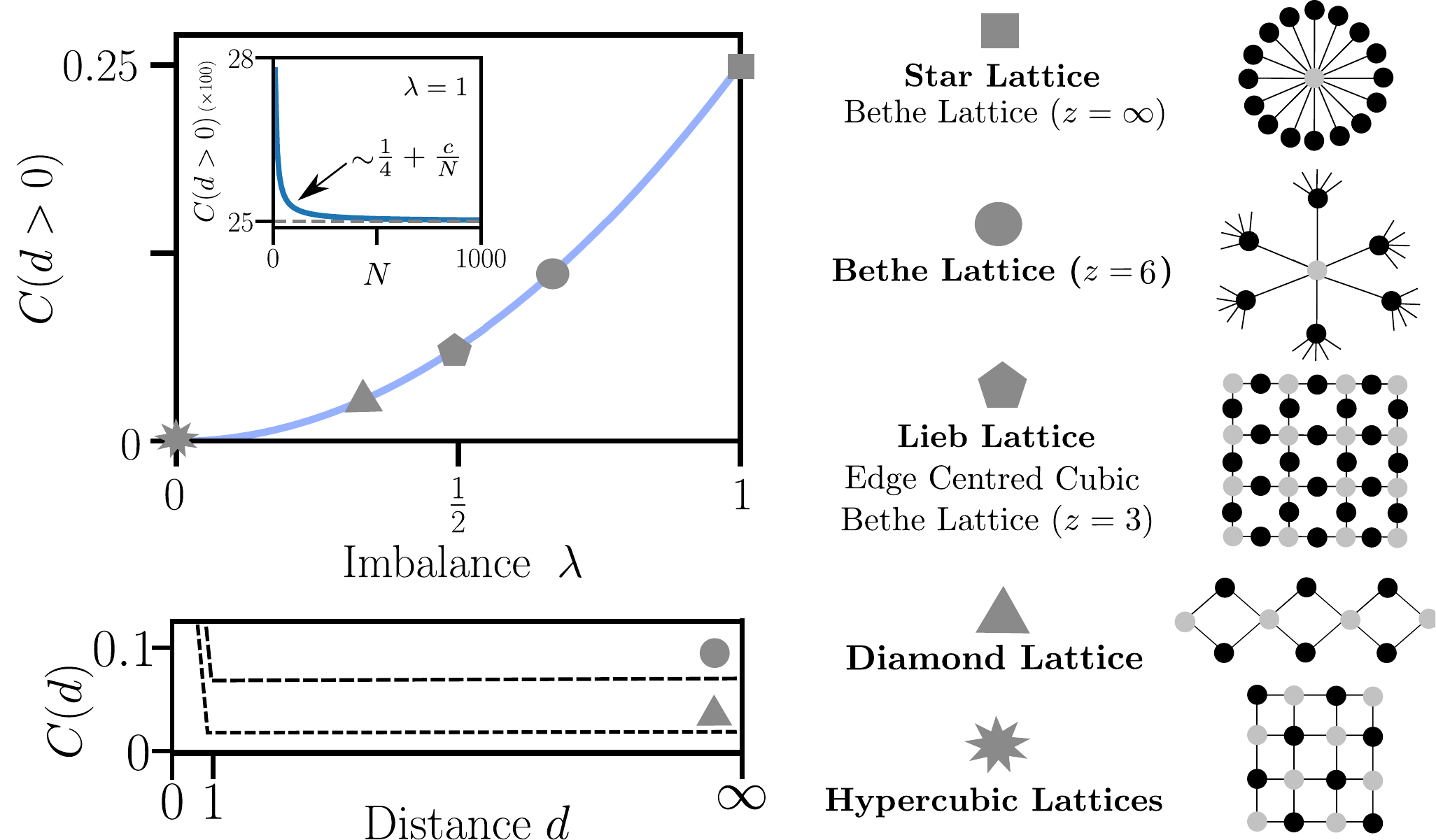}
\caption{Off-diagonal correlations versus lattice imbalance ratio $\lambda = (N_{A}-N_{B})/(N_{A} + N_{B})$ for the maximum entropy steady states of the half-filled Hubbard model in the limit of large system size $N$ (actual data is for $N = 10^{3}$). These states are formed from Floquet heating the ground state whilst preserving the relevant SU(2) symmetry. If the ground state is for $U > 0$, the spin SU(2) symmetry must be conserved by the heating and the correlation function $C(d)$ is defined in Eq. (\ref{Eq:Correlation Function}) with $O = S$. For $U < 0$, the $\eta$ SU(2) symmetry must be conserved by the heating and the correlation function is defined in the same way but with $O = \eta$. The symbols (from left to right) correspond to imbalances of $0$, $1/3$, $1/2$, $2/3$ and $1$ respectively --- examples of corresponding lattices are listed on the right, with those titled in bold pictured. For the instances of the Bethe lattice, $z$ is the corresponding co-ordination number. Inset) Scaling of $C(d>0)$ with $N$ for $\lambda =1$, where convergence is slowest but still at a rate of $1/N$ ($c$ is a constant). 
Bottom left)  Correlation function $C(d)$ versus distance $d$ for the two indicated imbalances.}
\label{Fig:F2}
\end{figure}

\par The quasi-1D unbalanced lattices used in Figs. \ref{Fig:F1} and \ref{Fig:F15} were chosen because they are amenable to Matrix Product based routines, allowing us to compare the steady state correlations to the ground state and perform the time-dynamics for system sizes larger than those of exact diagonalisation. Nonetheless, these results are still limited to finite-size lattices due to the computational expense of the many-body problem. Our findings, however, apply to unbalanced lattices of any size $N$ and any physical dimension $D$ and one of the key results of this paper is that the induced order will remain finite and not decay away to $0$ for increasing $N$. To access and quantify the steady state order in this limit we therefore utilise the analytical solution of Ref. \cite{HeatingInducedOrder3} to calculate the off-diagonal order for increasing system size $N$ until convergence is achieved. We find that a value of $N = 10^{3}$ is sufficient for our desired accuracy\footnote{For $N=10^{3}$ the area under the $C(d) > 0$ versus $\lambda$ curve - see Fig. \ref{Fig:F2} - changes by less than $0.1 \%$ when $N$ is doubled and thus the change cannot be visually discerned} as, for any given $\lambda$, the off-diagonal correlations converge to their asymptotic value \textit{at least} as fast as $1/N$. 
\par In order to be as general as possible, we consider \textit{any} bi-partite lattice in the limit of large $N$ and quantify the long-range order in the maximum entropy state, following Floquet heating of the ground state, as a function of the lattice `imbalance' \newline $\lambda = (N_{A}-N_{B})/(N_{A} + N_{B})$. The lattices $\Lambda_{1}$ and $\Lambda_{2}$ from Fig. \ref{Fig:F1} have $\lambda = 1/3$ and $\lambda = 1/2$ respectively whilst a 1D chain clearly has $\lambda = 0$. Figure \ref{Fig:F2} plots this order for the full range of $0 \leq \lambda \leq 1$ and we observe how the steady state order increases monotonically with the imbalance; only the maximum entropy states on \textit{balanced} lattices are incapable of hosting off-diagonal order for large $N$. We have also marked some notable lattices and their corresponding imbalances in this figure: these include the Lieb lattice and the edge-centred-cubic lattice, which are known to arise in materials such as the cuprates and perovskites \cite{CopperOxideLayers1, CopperOxideLayers2, PerovskiteLike1}.        

\par On any  unbalanced lattice in the limit of large $N$ with  $\lambda > 0$ the maximum entropy states satisfy Yang's definition for ODLRO \cite{Yang1} as we have that $\lim_{|a - b| \rightarrow \infty}\rho(a, b) = {\rm const} \neq 0$. The matrix $\rho(a, b)$ is the two-body reduced density matrix and corresponds to ${\rm Tr}(\rho_{\rm ss}S^{+}_{a}S^{-}_{b})$ for $U > 0$ and  ${\rm Tr}(\rho_{\rm ss}\eta^{+}_{a}\eta^{-}_{b})$ for $U < 0$. Whilst the ground state on an unbalanced lattice will also typically satisfy this ODLRO property there is a crucial difference. Specifically, unlike the steady state the two-body reduced density matrix for the ground state is not translationally invariant, i.e. $\rho(a, b)$ is not constant for all $|a - b| = d$, an immediate consequence of the fact the lattice is not spatially symmetric. 
\par The combination of particle-hole ODLRO and translational invariance in an equilibrium state (which these steady states are as they commute with $H$ in Eq. (\ref{Eq:Hamiltonian})) has been proven to imply both the Meissner effect and flux quantisation \cite{Meissner1, Meissner2}, suggesting that our steady states are superconducting for $U < 0$. Without translational invariance these proofs no longer hold and thus we cannot say the same of the ground state. The inhomogeneous nature of the correlations (the correlations are inhomogeneous even if we ignore the changing sign) may be suppressing the state's underlying order.

\section{Discussion - Experimental Realisation}
\par Whilst the Meissner effect and flux quantisation will only be observable for $U < 0$ and $D \geq 2$ we anticipate that other crucial differences will arise when probing the steady state vs ground state on unbalanced lattices in both the attractive and repulsive case. Ultracold quantum simulators offer an experimental setting in which the steady states in this paper can be realised, and such differences identified. 

\par The fermionic Hubbard model is realisable in this setting by loading an ultracold atomic gas into the potential landscape generated by standing wave laser beams \cite{UltracoldHubbard2}. The Hubbard interaction can be directly tuned to be both positive or negative via Feshbach resonances \cite{UltracoldHubbard1, FeshbachOpt}. The lattice geometry is dependent on the interference pattern the standing wave lasers create and an optical Lieb lattice has already been successfully realised and loaded with bosons \cite{OpticalLieb1, OpticalLieb2}. The setup necessary for creating the diamond lattice ($\Lambda_{2}$ in Fig. \ref{Fig:F1}) has also been theorised for both fermions and bosons \cite{OpticalDiamond1, OpticalDiamond2}. 
\par In order to realise the desired steady states the ground state must be driven out of equilibrium and undergo Floquet heating whilst preserving the desired SU(2) symmetry. Inducing a periodic modulation of either the hopping strength or the interaction strength will do this and can be achieved by periodically `shaking' the standing wave interference pattern \cite{Messer_Floquet_2018}. 
\par The timescale on which the desired Floquet heating occurs and the uniform correlations are induced will depend on the specifics of the lattice geometry, driving field and the parameters of $H$. Nonetheless, using a moderate interaction strength and resonantly matching the driving frequency to it should dramatically reduce this timescale \cite{ Timescales2}. Studies of Floquet heating rates for infinite Bethe lattices suggest such a timescale is on the order of tens of hopping times \cite{Timescales1, Timescales3}.    
\par Alongside direct measurement of two-point correlation functions\footnote{Correlations along the $x$ and $y$ spin axes can be measured directly via in-situ imaging \cite{SpinSpin1} whilst $\eta$ correlations could be measured by associating doublons to molecules and performing time-of-flight measurements \cite{EtaCorrMeasure}} which would verify the results presented here, measurements can be taken in this setup which cannot be done computationally. The optical conductivity spectrum provides information about the transport properties of a state and can be used to distinguish superconductors, conductors and insulators. This cannot be calculated computationally for the steady states in this paper as it requires access to the eigenspectrum of the Hubbard Hamiltonian on an unbalanced lattice for system sizes well beyond the reach of exact diagonalisation \cite{OpticalConductivityCalc1}. In optical lattices, however, this spectrum can be accessed via spectroscopic techniques \cite{OpticalConductivity1, OpticalConductivity2, OpticalConductivity3}. 
\par Ultracold quantum simulators thus offer the opportunity to realise the states uncovered in this paper and identify distinctive features, outside of the two-point correlation function, which set them apart from the ground state. 
\section{Conclusion and Outlook}
We have shown how, in the macroscopic limit, the maximum entropy steady states of the Hubbard model on unbalanced lattices possess finite, uniform ODLRO. These states can be formed by periodically driving the ground state of the system whilst preserving the SU(2) symmetries of the lattice; a process which renormalizes the staggered, inhomogeneous correlations initially present. For the attractive Hubbard model $U < 0$ the steady state is arguably a superconductor whilst the same arguments cannot be applied to the ground state. We quantified the induced steady state order as a function of the \textit{imbalance} of the lattice, allowing us to make predictions for \textit{any} bi-partite lattice. Finally, we discussed how such exotic, desirable steady states can be formed and probed with current technologies in optical lattice setups. 
\par Our results here provide an understanding of how driving can expose and manifest order which is `suppressed' in equilibrium due to the lattice geometry - independent of any microscopic parameters. This phenomenon should be observable beyond the bi-partite, single-band Hamiltonian that we have studied here; equilibrium states with the desired, inhomogeneous long-range order can arise in Hubbard models with more than two fermionic species \cite{SUNHubbard}, multiple-bands \cite{MultipleBands} and even away from the bi-partite regime \cite{TindallKappa}. Moreover, regardless of how such states arise, we know that driving which preserves the requisite symmetry is guaranteed to transform the state into one which manifests uniform ODLRO in the thermodynamic limit. 
\par It is worth stating, however, that in any realistic experimental setting there will be unwanted influences which break these symmetries to some degree. In the context of finite, balanced bi-partite lattices and non bi-partite lattices it has been shown that heating-induced order is still manifest on transient timescales if the symmetry breaking is sufficiently small \cite{HeatingInducedOrder1, Timescales2}. We believe that such results should hold in the context of the setups considered here and rigorously quantifying this transient timescale as a function of the lattice geometry, imbalance, driving parameters and system size will be an important aspect of future work.    
\par Finally, the emergence of superconducting order has recently been observed in a number of solid-state materials upon exposure to strong driving \cite{Organic1, Organic2, Organic3, Organic4, Organic7}. These materials often contain layers of compounds such as ${\rm Cu}{\rm O}_{2}$ which can be described with single or multi-orbital Hubbard Hamiltonians on unbalanced Lieb lattices \cite{LiebLike2, CopperOxideLayers1}. Whilst a microscopic description of these materials is beyond the scope of this paper, our results offer a possible phenomenological mechanism for these experimental observations. The optical conductivity spectrum is typically used as a fingerprint for the emergence of superconductivity in these solid-state experiments \cite{OpticalSCReview2}. Measurement of this spectrum via an optical lattice realisation of our steady states could strengthen the connection between our work and these experiments.     

\textit{Acknowledgements}
This work has been supported by EPSRC grants No. EP/P009565/1 and EP/K038311/1 and is partially funded by the European Research Council under the European Union’s
Seventh Framework Programme (FP7/2007-2013)/ERC Grant Agreement No. 319286 Q-MAC. JT is also supported by funding from Simon Harrison. MAS acknowledges support by the DFG through the Emmy Noether programme (SE 2558/2-1) and F. S. acknowledges support from the Cluster of Excellence `Advanced Imaging of Matter' of the Deutsche Forschungsgemeinschaft (DFG) - EXC 2056 - project ID 390715994.

\end{document}